\documentstyle{article}
\date{\today}

\begin{document}
\title{Electromagnetic Decay of Vector Mesons as derived from QCD Sum Rules}

\author{{Shi-lin Zhu,$^1$ W-Y. P. Hwang,$^{2,3}$ and Ze-sen Yang$^1$}\\
{$^1$Department of Physics, Peking University, Beijing, 100871, China}\\
{$^2$Department of Physics, National Taiwan University, Taipei, 
Taiwan 10764}\\
{$^3$Center for Theoretical Physics, Laboratory for Nuclear Science 
       and}\\
{Department of Physics, Massachusetts Institute of Technology, }\\ 
{Cambridge, Massachusetts 02139}
}
\maketitle

\begin{center}
\begin{minipage}{120mm}
\vskip 0.6in
\begin{center}{\bf Abstract}\end{center}
{We apply the method of QCD sum rules in the presence of the external 
electromagnetic $F_{\mu\nu}$ field to the problem of the electromagnetic 
decay of the various vector mesons, such as $\rho \to \pi+ \gamma$, $
K^\ast \to K + \gamma$, and $\eta'\to \rho +\gamma$. The induced 
condensates obtained previously from the study of baryon magnetic moments
are adopted, thereby ensuring the parameter-free nature of the present 
calculation. Further consistency is reinforced by invoking the various QCD 
sum rules for the meson masses. The numerical results on the various 
radiative decays agree very well with the experimental data. 

PACS Indices: 12.38.Lg; 13.40.Hq; 14.40.Aq
}
\end{minipage}
\end{center}

\large

\vskip 1.5cm

\par
Quantum chromodynamics (QCD) is believed to be the underlying theory 
of strong interactions. However, little is known concerning the structure of 
a meson (for which the naive quark-antiquark picture is at most only the 
leading approximation), since QCD is nonperturbative at the typical hadronic 
scale. In
the absence of the first principle calculation,  the electromagnetic decay
of a vector meson, such as $\rho \to \pi \gamma$, may be treated via an
effective Lagrangian \cite{VMD,LLLL}. In the vector meson dominance model (VDM)
\cite{VMD}, an effective Lagrangian with small flavor $SU(3)$ symmetry 
breaking was used while in chiral perturbation theory \cite{LLLL} the 
Bardeen-subtracted Wess-Zumino action, together with symmetry-breaking 
corrections $F_K/F_{\pi}\approx F_{\eta}/F_{\pi}\approx 1.2$ and a given
$\omega$-$\phi$ mixing parameter $\epsilon$, was employed to predict 
radiative vector-meson decays. As pointed out in \cite{LLLL}, VDM may work 
to the level of within $20\%$ for the flavor $SU(2)$ sector
but is expected to be considerably worse for flavor $SU(3)$. 
Another approach, which is the focus of the present paper, is made possible 
via the method of QCD sum rules \cite{SVZ} in which the nonperturbative
QCD physics is incorporated systematically as power corrections in the 
short-distance operator product expansion (OPE). Results from QCD sum rules
will certainly serve as another useful reference point concerning radiative 
decays of vector mesons.

\par
To be specific, we consider the electromagnetic decay of a vector meson in 
the flavor $SU(3)$ sector, i.e. in the sector of (u, d, s). Using $\rho \to
\pi +\gamma$ as the illustrative example, we may describe the decay process 
by the effective Lagrangian,
\begin{equation}
{\cal L}_{\rho\pi\gamma}=e\frac{g_{\rho\pi\gamma}}{m_\rho} 
\epsilon_{\mu\nu\lambda\delta}
\partial^{\nu}A^{\mu} \rho^{\lambda} \partial^{\delta} \pi \, ,
\end{equation}
where $A_{\mu}$ is the electromagnetic field. The corresponding partial width 
is
\begin{equation}
\Gamma(\rho\to\pi\gamma)=\alpha\frac{g^2_{\rho\pi\gamma}}{3} 
\frac{P_{\mbox{cm}}^3}{m_\rho^2}  \, ,
\end{equation}
with $P_{\mbox{cm}}$ is the pion momentum in the CM frame (i.e. in the rest frame of 
the vector meson).

\par
In the original work of Shifman, Vainshtein, and Zakharov \cite{SVZ}, the 
nonperturbative QCD effects are incorporated through the various non-zero 
condensates in the nontrivial QCD vacuum. Ioffe and Smilga \cite{IOFFE} 
and, independently, Balitsky and Yung \cite{Balit} studied baryon magnetic
moments by extending the method of QCD sum rules in the presence of 
the external electromagnetic field. It turns out that the QCD sum rule
method as obtained in \cite{IOFFE} or \cite{Balit} can readily be applied 
to the problem of the electromagnetic decay of the various vector mesons, 
such as $\rho \to \pi+ \gamma$, $K^\ast \to K + \gamma$, and $\eta'\to 
\rho +\gamma$. As a result, we wish to pursue along this direction a little
further, hoping to obtain results which may help to clarify certain issues in 
relation to the effective Lagrangian approach \cite{VMD,LLLL}.

\par
We recall \cite{IOFFE,Balit} that, in the method of QCD sum rules, the 
two-point correlation function $\Pi_\mu(p)$ in the presence of the constant 
electromagnetic field $F_{\mu\nu}$ may be introduced as follows,
\begin{equation}
\label{3}
\Pi_\mu (p)=i\int d^4x e^{ipx}\langle 0|T[j_{\rho_{\mu}}(x),j_{\pi}(0)]
|0\rangle_F \, ,
\end{equation}
where we introduce the interpolating fields,
\begin{equation}
\begin{array}{lll}
j_{\rho_{\mu}}(x)={\overline u}^a(x)\gamma_{\mu} d^a(x) \, , &   &
j_{\pi}(x)={\overline u}^a(x)i\gamma_5 d^a(x) \, ,
\end{array}
\end{equation}
with $a$ the color index. As in the QCD sum rule analyses for nucleon
properties, we may introduce the overlap amplitude (coupling) of the meson 
current to the (physical) meson:
\begin{equation}
\langle 0|j_{\rho_{\mu}}|\rho \rangle =\lambda_{\rho}\epsilon_{\mu} \, ,
\end{equation}
\begin{equation}
\langle 0|j_{\pi}|\pi \rangle =\lambda_{\pi} \, ,
\end{equation}
where $\epsilon_{\mu}$ is the polarization $4-$vector of the $\rho$ meson.

\par 
Embedding the system in the constant $F_{\mu\nu}$ field and introducing 
intermediate states we express the correlation function at the hadronic level
as follows,
\begin{equation}
\label{7}
\Pi_\mu (p)=
\lambda_{\rho} \lambda_{\pi} \frac{g_{\rho\pi\gamma}}{2m_{\rho}} \frac{1}
{p^2 -m_{\rho}^2}
\frac{1}{p^2 -m_{\pi}^2} \epsilon_{\alpha\beta\mu\sigma} F^{\alpha\beta} 
p^{\sigma}
+\Pi_\mu^{\rho^{\ast}\rightarrow\pi\gamma}(p)
+\cdots.
\end{equation}
The $\rho\pi\gamma$ coupling is defined through the relation:
\begin{equation}
\langle \pi^l(p^{\prime}) | j^{em}_{\lambda} |\rho^m_{\mu}(p) \rangle \equiv
-i\delta^{lm} e\frac{ g_{\rho\pi\gamma} }{m_{\rho}} 
\epsilon^{\lambda\nu\sigma\mu} p_{\nu} q_{\sigma} K(q^2),
\end{equation}
where $K(q^2)$ is the form factor normalized such that $K(0)=1$. The 
superscripts $l$ and $m$ are isospin indices. Note that, in Eq. (7),
we write down explicitly the leading $\rho\to\pi\gamma$ contribution, label 
the next transition term by $\Pi_\mu^{\rho^{\ast}\rightarrow\pi\gamma}(p)$, 
and denote the continuum contribution simply by the ellipsis.

\par
The external field $F_{\mu\nu}$ may induce changes in the physical vacuum and 
modify the propagation of quarks. Up to dimension six ($d\leq 6$), we may
introduce three induced condensates as follows.
\begin{equation}
\langle 0 | {\overline  q} \sigma_{\mu\nu} q |0 \rangle_{F_{\mu\nu}} = e_q e \chi
F_{\mu\nu} \langle 0 | {\overline  q}  q |0 \rangle,
\end{equation}
\begin{equation}
g_c \langle 0 | {\overline  q} \frac{\lambda^n}{2}G^n_{\mu\nu} q |0 \rangle_{F_{\mu\nu}}
= e_q e \kappa F_{\mu\nu} 
\langle 0 | {\overline  q}  q |0 \rangle,
\end{equation}
\begin{equation}
g_c \epsilon_{\mu\nu\lambda\sigma} 
\langle 0 | {\overline  q} \gamma_5 \frac{\lambda^n}{2} G^n_{\lambda\sigma} q |0 \rangle_{F_{\mu\nu}}
= i e_q e \xi F_{\mu\nu} 
\langle 0 | {\overline  q}  q |0 \rangle \, ,
\end{equation}
where $q$ refers mainly to $u$ or $d$ (with suitable modifications for the $s$ 
quark), $e$ is the unit charge, $e_u=\frac{2}{3}$, and $e_d=-\frac{1}{3}$. In
studying baryon magnetic moments, Ioffe and Smilga \cite{IOFFE} obtained the 
nucleon anomalous magnetic moments $\mu_p =3.0$ and  
$\mu_n=-2.0$ $(\pm 10\%)$ with the quark condensate susceptibilities 
$\chi \approx -8\,\mbox{GeV}^2$ and $\kappa$, $\xi$ quite small. 
Balitsky and Yung \cite{Balit} adopted the one-pole approximation to estimate
the susceptibilities and obtained,
\begin{equation}
\begin{array}{lll}
\chi=-3.3\,\mbox{GeV}^{-2},& \kappa=0.22,& \xi=-0.44 \, .
\end{array}
\end{equation}
Subsequently Belyaev and Kogan \cite{Belyaev} extended the calculation and 
obtained an improved estimate $\chi =-5.7\,\mbox{GeV}^{-2}$ using the two-pole 
approximation. Chiu et al \cite{Chiu} also estimated the susceptibilities 
with the two-pole model and obtained
\begin{equation}
\begin{array}{lll}
\chi=-4.4\,\mbox{GeV}^{-2},& \kappa=0.4,& \xi=-0.8 \, .
\end{array}
\end{equation}
Furthermore, Chiu and co-workers \cite{Chiu} re-analysed the various sum 
rules by treating $\chi$, $\kappa$, and $\xi$ as free parameters to provide
an overall fit to the observed baryon magnetic moments.
The opitimal values which they obtained are given by 
\begin{equation}
\begin{array}{lll}
\chi=-3\,\mbox{GeV}^{-2},& \kappa=0.75,& \xi=-1.5 \, .
\end{array}
\end{equation}
We note that in these analyses the susceptibility values are consistent 
with one another except the earliest result $\chi=-8\,\mbox{GeV}^{-2}$ 
in \cite{IOFFE}, which is considerably larger (in magnitude). In what 
follows, we shall adopt the condensate parameters $\chi=-3.5\,
\mbox{GeV}^{-2}$, 
$2\kappa +\xi \approx 0$ with $\kappa$, $\xi$ quite small, which represent the 
average of the latter values discussed above.

\par
On the other hand, the correlation function $\Pi_\mu(p)$ may also be 
evaluated at the quark level. 
\begin{equation}
\Pi_\mu (p)=-i \int d^4x e^{ipx}
{\bf Tr}[iS_u^{ab}(x) \gamma_{\mu} iS_d^{ba}(-x) i\gamma_5 ],
\end{equation}
where $iS^{ab}(x)$ is the coordinate-space propagator in the presence of 
the $F_{\mu\nu}$ field \cite{IOFFE,Balit}.

\par
Equating the correlation function obtained both at the quark level (l.h.s.)
and at the hadron level (r.h.s.), performing the Borel transform on both 
the sides (to accentuate the contribution from the leading $\rho\to\pi\gamma$
contribution), we finally obtain the QCD sum rule for $g_{\rho\pi\gamma}$:
\begin{equation}
\label{final}
\begin{array}{ll}
(e_u+e_d){a_q}{1\over 24\pi^2}
&[-3\chi +(2+ 2\kappa +\xi )\frac{1}{M^2_B}]+\cdots \\
&=\frac{g_{\rho\pi\gamma}}{2m_{\rho}} 
\frac{\lambda_{\rho}\lambda_{\pi}}{m_{\rho}^2 -m^2_{\pi}} 
(e^{-\frac{m_{\pi}^2}{M^2_B}}-e^{-\frac{m_{\rho}^2}{M^2_B}}) 
+\Pi_{\rho^{\ast}\rightarrow \pi\gamma}(M),
\end{array}
\end{equation}
where $a_q \equiv -(2\pi)^2 \langle 0|{\overline q} q |0\rangle$ and $M_B$ is 
the Borel mass (with $M^2_B$ playing essentially the role of $p^2$).
Note that in this sum rule (16) the dominant contribution comes from 
the induced quark condensate characterized by the susceptibility $\chi$. 
The contribution from $\rho^*\to\pi\gamma$ is suppressed 
by the factor $(\frac{m_{\rho}}{m_{\rho^{\ast}}})^3 $ and so is negligible 
to within $10\%$ (which is the typical accuracy of the QCD sum rule approach).
The sum rule (\ref{final}) is stable within the working interval 
$M_B^2\sim m_{\rho}^2$. 

\par
We proceed to note that in the framework of QCD sum rules the overlap 
amplitude $\lambda_{\rho}$ can be determined in a self-consistent manner 
by making use of the mass sum rules for the $\rho$ and $\pi$ 
mesons\cite{SVZ,REINDERS},
\begin{equation}
\lambda_{\rho}^2 =\frac{1}{\pi^2} e^{\frac{m_{\rho}^2}{M^2_B}} \{ \frac{1}{4}
 (1+\frac{\alpha_s}{\pi}) M^4_B E_1 -\frac{b}{48} +
\frac{\alpha_s}{\pi}\frac{14}{81} a^2_q  \frac{1}{M^2_B} \},
\end{equation}
\begin{equation}
\label{pion}
\lambda_{\pi}^2 =\frac{1}{\pi^2} e^{\frac{m_{\pi}^2}{M^2_B}} \{ \frac{3}{8} (1+
\frac{11}{3}\frac{\alpha_s}{\pi}) M^4_B E_1 +\frac{b}{32} +
\frac{\alpha_s}{\pi}\frac{7}{27} a^2_q  \frac{1}{M^2_B} \},
\end{equation}
where $E_1\equiv 1-(1+\frac{S_0^2}{M^2_B})e^{-\frac{S_0^2}{M^2_B}}$ (with
$S_0^2=1.5$GeV$^2$) is the factor arising from a quark-level approximation 
for the continuum, and $b\equiv (2\pi)^2\langle 0|\frac{\alpha_s}{\pi}
G^n_{\alpha\beta}G^n_{\alpha\beta}|0\rangle $ is the gluon condensate. 
It was pointed out in \cite{Novikov} that the instanton contributions may 
invalidate the usual sum rule techniques for the pseudoscalar current. 
Nevertheless, it was suggested in \cite{SVZ,REINDERS} that Eq. (18) may 
still provide a rough estimate of $\lambda_{\pi}$ with the experimental 
pion mass as input (despite the fact that it cannot
be used to predict the pion mass). Moreover there are many careful analyses of 
this correlator which yield very reasonable results for $\lambda_{\pi}$ 
\cite{Becchi,Narison,DOSCH}. We therefore adopt the value from the 
QCD sum rule for $\lambda_{\pi}$ with the physical mass $m_{\pi}$ as the 
input. We obtain the estimates $\lambda_{\pi}\approx 0.17\pm 0.03\, 
{\mbox{GeV}}^{2}$ and $\lambda_{\rho}\approx 0.17\,{\mbox{GeV}}^{2}$. 
An independent phenomenological analysis in \cite{SHURYAK} 
yields $\lambda_{\rho}= f_{\rho} m_{\rho}=0.17{\mbox{GeV}}^{2}$,
$\lambda_{\pi}=f_{\pi}\frac{m_{\pi}^2}{m_u +m_d}=0.20\pm 0.04
{\mbox{GeV}}^{2}$, values consistent with those from the present QCD sum rule
approach. 

Using our result on $\lambda_\rho$ and $\lambda_\pi$, the observed masses, 
and the condensate parameters discussed earlier, we finally obtain from the
QCD sum rule (16),
\begin{equation}
g_{\rho\pi\gamma}=0.59 \, ,
\end{equation}
which is very close to the value $g_{\rho\pi\gamma}=0.58$ as may be 
extracted from the partial width for the process $\rho\rightarrow \pi\gamma$.

\par
It is clear that the method described so far for $\rho\to\pi\gamma$ can be 
generalized in a straightforward manner to other electromagnetic decays of
vector mesons, among which we have considered additional processes as listed
in Table I. The working interval for the Borel mass for the corresponding 
sum rules is $M_B^2\sim m_{par}^2$, where $m_{par}$ is the mass of the parent 
particle in the decay process. In the case for the meson containing a strange 
quark or antiquark, we have taken into account the contributions due to the 
nonzero quark mass $m_s \approx 150 \mbox{MeV}$, which in the cases such as  $K^{\ast} 
\rightarrow K\gamma $ may give rise to corrections of order of about $ 20\%$.
We also use $\langle 0| {\overline s} s|0\rangle \approx 0.8 \times 
\langle 0|{\overline u} u |0\rangle$ (as suggested by chiral perturbation 
theory) which is essential for the $K^{\ast} \rightarrow K\gamma $ channel.    

\par
In Table I, we present the experimental values for the various electromagnetic 
decays of vector mesons, the predictions from VDM \cite{VMD}, those from the 
effective Lagrangian approach \cite{LLLL}, and what we have obtained from 
the QCD sum rules (this work). The results 
shown in this table indicate that the QCD sum rules yield predictions which are 
in better overall agreement with the experimantal values.

\vskip 1 true cm
\par
\noindent
TABLE I. Comparison between experimental data \cite{DATA} and predictions 
from the Vector Meson Dominance Model (VDM) \cite{VMD}, effective chiral 
Lagrangian approach  \cite{LLLL}, and QCD sum rules. The unit is KeV.

\vskip 0.5 true cm
\begin{tabular}{|c|c|c|c|c|}
\hline
 & experimental  & Vector Meson & effective chiral & QCD sum  \\
 & value & Dominance & Lagrangian & rules \\ 
\hline
 $\omega\rightarrow\pi\gamma $ &$716$&$888.0$&$800$& $720$ \\ 
 $\omega \rightarrow \eta \gamma $&$7$&$5.4$&$5$&$7.1$  \\ 
 $\rho\rightarrow \pi \gamma $&$68$&$63.0$&$80$&$65$  \\ 
 $\rho \rightarrow \eta \gamma $&$58$&$55.0$&$38$&$48$  \\ 
 $\phi\rightarrow \eta \gamma $&$57$&$71.0$&$68$& $80$ \\ 
 $\phi\rightarrow \eta^{\prime} \gamma $&$<1.8$&$0.24$&$1$&$0.67$  \\ 
 ${K^0}^{\ast}\rightarrow K^0  \gamma $&$116$&$149.0$&$117$& $102$ \\ 
 ${K^+}^{\ast}\rightarrow K^+ \gamma $&$50$&$68.0$&$29$& $60$ \\ 
 $\eta^{\prime}\rightarrow \rho \gamma $&$60$&$119.0$&$77$& $60^a$ \\ 
 $\eta^{\prime}\rightarrow \omega \gamma $&$6$&$12.5$&$7$& $6.1$  \\ 
 $\phi\rightarrow \pi \gamma $&$5.8$&$6.1$&$5$&$5.8^b$  \\
 \hline
\end{tabular}
\vskip 0.5 true cm
\noindent
$^a$ Used as the input to determine $\eta$-$\eta^{\prime}$ mixing 
angle $\theta$.\\
$^b$ Used as the input to determine $\phi$-$\omega$ mixing parameter 
$\epsilon$. \\
\vskip 1.0 true cm

\par
We note that all parameters have been determined consistently in the 
framework of QCD sum rules as illustrated before. In particular, the various 
overlap amplitudes (couplings) are determined from the various mass sum rules:
$\lambda_{\pi}=0.17\pm 0.03 \mbox{GeV}^{-2}$, 
$\lambda_K=0.23\pm 0.03 \mbox{GeV}^{-2}$, 
$\lambda_{\rho}=0.17\pm 0.01 \mbox{GeV}^{-2}$, 
$\lambda_{K^{\ast}}=0.21\pm 0.02 \mbox{GeV}^{-2}$. 
The above predictions and the other overlap amplitudes given in the following 
after taking into account the mixing effects are in good agreement with those 
in \cite{SHURYAK}. Note that the value for $\lambda_{\eta^{\prime}}$ is of 
some specific interest in view of the $U_A(1)$ problem.

\par
In the case of the $\phi$-$\omega$ mixing, we introduce 
\begin{equation}
\begin{array}{ll}
\label{22}
|\omega\rangle =\cos\epsilon |\omega_0\rangle -\sin\epsilon |\phi_0\rangle , &
|\phi\rangle =\sin\epsilon |\omega_0\rangle +\cos\epsilon |\phi_0\rangle  \, ,
\end{array}
\end{equation}
and
\begin{equation}
\begin{array}{ll}
\label{23}
j_{\omega} =\cos\epsilon\, j_{\omega_0} -\sin\epsilon\, j_{\phi_0} , &
j_{\phi} =\sin\epsilon\, j_{\omega_0} +\cos\epsilon\, j_{\phi_0} \, . 
\end{array}
\end{equation}
where 
\begin{equation}
\begin{array}{ll}
|\omega_0\rangle=\frac{1}{\sqrt{2}}(|u{\overline u}\rangle +|d{\overline d}\rangle), &
|\phi_0\rangle=|s{\overline s}\rangle \, .
\end{array}
\end{equation}
Using the interpolating field in (21), we derive the QCD sum rules for the
relevant radiative decays and the mass sum rules for $\omega$ and $\phi$ mesons.
We then use the observed value for the width of the decay $\phi \to\pi\gamma$ 
to determine the mixing angle. We obtain the mixing parameter 
$\epsilon =0.063\pm 0.005$ and the overlap amplitudes
$\lambda_{\omega}=0.16\pm 0.01 \mbox{GeV}^{-2}$ and $\lambda_{\phi}=0.24\pm 
0.02 \mbox{GeV}^{-2}$. As the Borel mass is varied in the interval 
$0.6\mbox{GeV}^2 \leq M^2_B \leq 1.1\mbox{GeV}^2$, the mixing parameter 
$\epsilon$ changes from $0.068$ to $0.058$. Note that, as a consistency 
check, our result is consistent with the value 
$\epsilon =0.079$ as determined from $\phi\rightarrow \rho\pi$ \cite{LLLL}. 

\par
Furthermore, we use the experimental partial width for the decay process 
$\eta^{\prime}\to \rho\gamma$ as the input in the derived QCD sum rules 
to extract the $\eta_0-\eta_8$ mixing angle, which is parametrized  
as in \cite{Donoghue,DATA}: 
\begin{equation}
\begin{array}{ll}
|\eta\rangle =\cos\theta |\eta_8\rangle -\sin\theta |\eta_0\rangle , &
|\eta^{\prime}\rangle =\sin\theta |\eta_8\rangle +\cos\theta |\eta_0\rangle  \, ,
\end{array}
\end{equation}
and correspondingly for the interpolating fields,
\begin{equation}
\label{25}
\begin{array}{ll}
j_{\eta} =\cos\theta j_{\eta_8} -\sin\theta j_{\eta_0} , &
j_{\eta^{\prime}} =\sin\theta j_{\eta_8} +\cos\theta j_{\eta_0} \, . 
\end{array}
\end{equation}
At the same time we obtain the $\eta$, $\eta^\prime$ mass sum rules with 
the interpolating fields in (24) and use the physical meson masses to 
calculate the overlap amplitudes. We then use the extracted mixing angle to 
predict other processes. 
Our results are $\lambda_{\eta}=0.23\pm 0.03 \mbox{GeV}^{-2}$, 
$\lambda_{\eta^{\prime}}=0.33\pm 0.05 \mbox{GeV}^{-2}$.  
$\theta =-19^\circ\pm 2^\circ$, a value consistent with that extracted from the decays
$\eta \rightarrow \gamma \gamma$ and $\eta^{\prime} \rightarrow \gamma 
\gamma$ \cite{Donoghue} and in \cite{DATA}. 
We find that such value on the $\eta_0 -\eta_8$ mixing is essential for the 
overall agreement with the experimental data. 

\par
We wish to emphasize the parameter-free nature of the present calculation, 
in that all parameters are determined consistently in the framework of QCD 
sum rules, with a single set of the values for the induced condensates 
obtained previously from studies of baryon magnetic moments. 
We note that the uncertainty of the overlap amplitudes (couplings) derived in 
this way might be as large as $10\%$ in some cases and the method of QCD sum 
rules, being an operator product expansion used for moderate $Q^2$, may 
have some inherent error of order of $10\%$. With this in mind, we may 
conclude that our results as shown in Table I agree very well with the 
experimental data. 

\par
This work was supported in part by the National Natural Science Foundation
of China and by Doctoral Program Foundation of the State Commission. It is
also supported in part by the National Science Council of R.O.C. (Taiwan)
under the grant NSC84-2112-M002-021Y.


\begin{thebibliography}{99}
\bibitem{VMD}J.W. Durso, Phys. Lett. B{\bf 184}, 348 (1987).
\bibitem{LLLL}U.-G. Meissner, Phys. Rep. {\bf 161}, 215 (1988).
\bibitem{SVZ}M.A. Shifman, A.I. Vainshtein, and V.I. Zakharov, Nucl. Phys. B
{\bf 147}, 385, 448 (1979).
\bibitem{IOFFE}B.L. Ioffe, and A.V. Smilga, Nucl. Phys. B{\bf 232}, 109 (1984).
\bibitem{Balit}I.I. Balitsky, and A.V. Yung, Phys. Lett. B{\bf 129}, 328 
(1983).
\bibitem{Belyaev}V.M. Belyaev, and Ya.I. Kogan, Yad. Fiz {\bf 40}, 1035 (1984)
[Sov. J. Nucl. Phys. {\bf 40}, 659 (1984).]
\bibitem{Chiu}C.B. Chiu, J. Pasupathy, and S.J. Wilson, Phys. Rev. D{\bf 33},
1961 (1986); C.B. Chiu, S.L. Wilson, J. Pasupathy, and J.P. Singh, Phys. 
Rev. D{\bf 36}, 1533 (1987).
\bibitem{REINDERS}L.J. Reinders, H. Rubinstein, and S. Yazaki, Phys. Rep. 
{\bf 127}, 1 (1985).
\bibitem{DATA}Particle Data Group, Phys. Rev. D{\bf 50}, 1173 (1994).
\bibitem{Novikov}V.A. Novikov, M.A. Shifman, A.I. Vainshtein, and V.I. 
Zakharov, Nucl. Phys. B{\bf 181}, 301 (1981).
\bibitem{Becchi}C. Becchi, S. Narison, E.de Rafael, and F.J. Yndurain, 
Z. Phys. C{\bf 8}, 335 (1981).
\bibitem{Narison}S. Narison, Riv. Nuovo. Cimento {\bf 10}, 2 (1987) and 
references therein.
\bibitem{DOSCH}H.G. Dosch, M. Jamin, and B. Stech, Z. Phys. C{\bf 42}, 167 
(1989).
\bibitem{SHURYAK}E.V. Shuryak, Rev. Mod. Phys. {\bf 65}, 1 (1993).
\bibitem{Donoghue}J.F. Donoghue, E. Golowich, and B.R. Holstein, {\sl 
Dynamics of the Standard Model, pp204} (Cambridge University Press, NewYork, 
1992) and references therein.
\end{thebibliography}
\end{document}